\documentclass[conference]{IEEEtran}
\IEEEoverridecommandlockouts

\usepackage{cite}
\usepackage{amsmath,amssymb,amsfonts}
\usepackage{algorithmic}
\usepackage{graphicx}
\usepackage{textcomp}

\usepackage{setspace,url}
\usepackage{numprint}
\usepackage{colortbl,dcolumn}
\npthousandsep{,}
\usepackage[dvipsnames]{xcolor}
\usepackage{makecell}
\usepackage{multirow}
\usepackage{cite}

\def\BibTeX{{\rm B\kern-.05em{\sc i\kern-.025em b}\kern-.08em
    T\kern-.1667em\lower.7ex\hbox{E}\kern-.125emX}}

\RequirePackage{etoolbox}
\RequirePackage{pgf} 
\RequirePackage{xcolor}

\newcommand{\gradientcelld}[8]{
\xdef\lowvalx{#2}%
\xdef\midvalx{#3}%
\xdef\maxvalx{#4}%
\xdef\lowcolx{#5}%
\xdef\midcolx{#6}%
\xdef\highcolx{#7}%
\xdef\opacityx{#8}%

\ifdimcomp{#1pt}{>}{\maxvalx pt}{\cellcolor{\highcolx!100.0!\midcolx!\opacityx}#1}{
\ifdimcomp{#1pt}{<}{\midvalx pt}{%
\ifdimcomp{#1pt}{<}{\lowvalx pt}{\cellcolor{\midcolx!0.0!\lowcolx!\opacityx}#1}{
     \pgfmathparse{int(round(100*(#1/(\midvalx-\lowvalx))-(\lowvalx*(100/(\midvalx-\lowvalx)))))}%
    \xdef\tempa{\pgfmathresult}%
    \cellcolor{\midcolx!\tempa!\lowcolx!\opacityx}#1%
}}{
     \pgfmathparse{int(round(100*(#1/(\maxvalx-\midvalx))-(\midvalx*(100/(\maxvalx-\midvalx)))))}
    \xdef\tempb{\pgfmathresult}%
    \cellcolor{\highcolx!\tempb!\midcolx!\opacityx}#1%
}}
}

\def\ninept{\def\baselinestretch{1.144}\let\normalsize\small\normalsize}

\begin{document}

\ninept

\title{Analysis of Speech Temporal Dynamics in the Context of Speaker Verification and Voice Anonymization\\
\thanks{This work was supported by the French National Research Agency under project Speech Privacy and project IPoP of the Cybersecurity PEPR.  Experiments were carried out using the Grid’5000 testbed.}
}


\author{
\IEEEauthorblockN{Natalia Tomashenko}
\IEEEauthorblockA{\textit{Universit\'e de Lorraine, CNRS, Inria} \\ \textit{Loria, F-54000} \\ Nancy, France \\
natalia.tomashenko@inria.fr}
\and
\IEEEauthorblockN{Emmanuel Vincent}
\IEEEauthorblockA{\textit{Universit\'e de Lorraine, CNRS, Inria} \\ \textit{Loria, F-54000 } \\ Nancy, France \\
emmanuel.vincent@inria.fr}
\and
\IEEEauthorblockN{Marc Tommasi}
\IEEEauthorblockA{\textit{Universit\'e de Lille, CNRS, Inria}  \\ \textit{Centrale Lille, UMR 9189 - CRIStAL} \\
Lille, France \\
marc.tommasi@inria.fr}
}

\maketitle

\begin{abstract}
In this paper, we investigate the impact of speech temporal dynamics  in application to automatic speaker verification and speaker voice anonymization tasks. 
We propose several metrics to perform automatic speaker verification based only on phoneme durations. Experimental results  demonstrate that phoneme durations leak some speaker information and can reveal  speaker identity from both original and anonymized speech. Thus, this work emphasizes the importance of taking into account the speaker's speech rate and, more importantly, the speaker's phonetic duration characteristics,  as well as the need to modify them in order to develop anonymization systems with strong privacy protection capacity.
\end{abstract}


\section{Introduction}
\label{sec:intro}

Speech data carries personal or sensitive information via speaker traits (e.g., identity, gender, age, ethnicity, accent), 
and sometimes via linguistic content (e.g., name, address)
and paralinguistic content (e.g., emotion).  
Most voice-based human-computer interaction technologies today rely on cloud-based machine learning systems trained on speech data collected from the users.
This poses serious privacy
risks and requires implementation of privacy-enhancing technologies to protect users' sensitive and private information.

One common approach to privacy protection of speech data is voice anonymization, which aims to suppress personally identifiable speaker traits, leaving linguistic and paralinguistic content intact \cite{tomashenko2020introducing}. 
Voice anonymization methods can be broadly classified into two categories. Signal processing based methods rely on simple signal transformations such as spectral warping using the McAdams coefficient \cite{patino2020speaker}, pitch shifting based on time-scale modification \cite{mawalim22_spsc}, and others \cite{gupta2020designn,tavi2022improving}.
By contrast, neural voice conversion based methods \cite{srivastava2021,miao2023speaker,champion2023,yao2024musa} rely on disentangling attributes such as content, speaker, pitch, emotion, etc., anonymizing the selected attributes, and generating the anonymized speech signal using a speech synthesis model. Most state-of-the-art voice conversion based anonymization methods use large-scale pre-trained models for extracting specific attributes and provide better content and privacy preservation than signal processing based methods.
The diversity of approaches is illustrated by the VoicePrivacy 2024 Challenge \cite{tomashenko2024voiceprivacy}, which provided six baseline anonymization systems, namely
anonymization using x-vectors and a neural source-filter model \cite{fang2019speaker,srivastava2021}, 
signal processing based anonymization using the McAdams coefficient \cite{patino2020speaker}, anonymization using  phonetic transcription and generation of artificial pseudo-speaker embeddings by a generative adversarial network (GAN) 
\cite{meyer2023prosody}
anonymization using neural audio codec (NAC) language modeling \cite{panariello_speaker_2023}, and
anonymization using acoustic vector quantization bottleneck (VQ-BN) features from an automatic speech recognition (ASR) acoustic model. 

While specific studies have been dedicated to speaker information carried by pitch \cite{tavi2022improving,srivastava2021,champion2023}, the impact of speech temporal dynamics on speaker verification and re-identification has been overlooked. As a result, all the baseline systems of the VoicePrivacy 2024 Challenge modify various characteristics of the input (original) speech signal linked with speaker identity
but keep speech rate and phoneme durations unchanged. Most other state-of-the-art anonymization systems also do not modify phoneme durations \cite{yao2024musa,miao2022language,miao2023speaker}.

Among the rare exceptions are cascaded ASR and text-to-speech (TTS) systems where word-level \cite{sinha2022eli} or phoneme-level transcripts obtained by an ASR system are provided to a TTS system for synthesis of the given linguistic content with a new target voice.
It can be assumed that these systems do not retain any information about speaker identity, however
they fail to preserve any paralinguistic attributes, which are required in real-life voice anonymization scenarios.
The most relevant work for our study is \cite{prajapati2022voice} where speed perturbation with a constant factor was used as an anonymization method either alone or in combination with anonymization based on a cycle consistent generative adversarial network (CycleGAN). The authors showed that speech rate perturbation with a constant factor degrades the performance of the automatic speaker verification systems associated with \textit{ignorant} and
\textit{lazy-informed} attackers~\cite{Tomashenko2021CSl}, but they did not consider the stronger \textit{semi-informed} attack model that is today's standard \cite{tomashenko2024first}.

Speed perturbation alone cannot be considered as a strong privacy protection method since a lot of speaker information remains in the speech signal after it. Yet we show that it is an essential ingredient in suppressing speaker information and must be taken into account in state-of-the-art anonymization systems. Indeed, while state-of-the art automatic speaker verification (ASV) systems do not explicitly rely on speaker temporal dynamics \cite{desplanques2020ecapa}, a few past studies have shown the applicability of durational characteristics for this task
\cite{bulgakova2015speaker,fujita21_interspeech,fujita2024speech}. 
Works \cite{fujita21_interspeech,fujita2024speech} propose to use speech rhythm-based emebeddings for speech synthesis.
We also show that more information about the speaker is contained in the temporal dynamics and the duration of phonemes than in the speech rate.

Our contributions build upon \cite{bulgakova2015speaker,fujita21_interspeech,fujita2024speech} and include: 
phoneme duration features and distance metrics for ASV based on phoneme durations (Section~\ref{sec:asv_using_phone_dur}); and
experimental evaluation and analysis of the resulting ASV performance on original data and data anonymized using two state-of-the-art anonymization systems with and without temporal dynamics modification (Section~\ref{sec:exprimental results}). 
To our best knowledge, this is the first work that performs such analysis and evaluation of the impact of speaker temporal dynamics on the anonymization task 
and demonstrates its importance for the design of voice anonymization systems.

\section{Speaker verification using phoneme duration dynamics}
\label{sec:asv_using_phone_dur}

\subsection{Metrics}
\label{sec:metrics}

We define two metrics to quantify the distance between speakers' temporal dynamics in the context of speaker verification. 
Let us denote by $N$ the number of phoneme classes $ph_1,\ldots,ph_N$, and
for two speakers $s_i$ and $s_j$ in the dataset
by $u_i^{1}\ldots u_i^{M_i}$ the utterances of speaker $s_i$ and
$u_j^{1}\ldots u_j^{M_j}$ the utterances of speaker $s_j$.

The first metric is based on the cosine distance between two vectors of mean phoneme durations:
\begin{equation}
\rho_1(s_i,s_j)=1-\cos(\boldsymbol{\mu_{i}},\boldsymbol{\mu_{j}}),
\end{equation}
where $\boldsymbol{\mu_{i}}$,
$\boldsymbol{\mu_{j}}$ are $N$-dimensional vectors composed of the average lengths of phonemes $ph_1,\ldots,ph_N$ computed over utterances $u_i^{1}\ldots u_i^{M}$ 
  (for ${\boldsymbol{\mu}}_{i}=
[\mu_{i}^{\left(1\right)},\ldots,\mu_{i}^{\left(N\right)}]$
) 
and 
$u_j^{1}\ldots u_j^{M}$ (for ${\boldsymbol{\mu_{j}}}=
[\mu_{j}^{\left(1\right)},\ldots,\mu_{j}^{\left(N\right)}]$),
respectively.
In case phoneme $ph_k$ is missing  in the considered utterances or has less then 
a given number of 
instances (considered as a threshold parameter), its mean values in $\boldsymbol{\mu_{i}}$ or $\boldsymbol{\mu_{j}}$ are replaced by the global mean duration of all phonemes in the  considered utterances for a given speaker.
Before computing metric $\rho_1$, mean normalization is applied to all $\boldsymbol{\mu_{i}}$ and $\boldsymbol{\mu_{j}}$.

We also propose a second metric that is defined as follows:

\begin{equation}
\rho_2(s_i,s_j)=1-\frac{1}{N}\sum_{k=1}^{N}\min\left\{\frac{\mu_i^{(k)}}{\mu_j^{(k)}}, \frac{\mu_j^{(k)}}{\mu_i^{(k)}}\right\}.
\end{equation}

Based on the proposed metrics we can perform ASV and compute an equal error rate (EER). Such ASV systems can also be considered as attackers for the anonymization task.

\subsection{Phoneme sets}
\label{sec:phoneme_set}
We experimented with two sets of phonetic classes: (1) $N=39$ phonemes based on the ARPAbet symbol set corresponding to the Carnegie Mellon University pronunciation dictionary\footnote{\url{http://www.speech.cs.cmu.edu/cgi-bin/cmudict}}, not counting variations due to lexical stress; and (2) $N=336$ phoneme classes that take into account position in the word and stress.

\section{Experimental results}
\label{sec:exprimental results}

\subsection{Data}
\label{sec:data}

Experiments were conducted on the \textit{LibriSpeech}\footnote{\label{fn:url1}\textit{LibriSpeech}: \url{http://www.openslr.org/12}}~\cite{panayotov2015librispeech} corpus of read English audiobooks, which was used in all VoicePrivacy Challenge editions. It contains approximately \numprint{1000}~hours of speech from \numprint{2484} speakers sampled at 16~kHz.
We conducted the first series of experiments and analyses on the full \textit{LibriSpeech-train-960} dataset that contains data from \numprint{2338} speakers.
In the second series of experiments, we used the \textit{LibriSpeech-train-clean-360} subset anonymized by the two different speaker voice anonymization systems denoted SAS-1 and SAS-2 described in Section~\ref{sec:anonym_systems}.
To perform phoneme segmentation, four triphone Gaussian mixture model - hidden Markov model (GMM-HMM) acoustic models were trained using the Kaldi speech recognition toolkit \cite{povey2011kaldi} on the following training data: (1)~original   \textit{LibriSpeech-train-960}; 
    (2)~original   \textit{LibriSpeech-train-clean-360}; 
    (3)~\textit{LibriSpeech-train-clean-360} anonymized by SAS-1; 
    and 
    (4)~\textit{LibriSpeech-train-clean-360} anonymized by SAS-2.
Statistics for the number of trials used in ASV evaluation are given in Table~\ref{tab:trials}.

  
\begin{table}[ht!]
\centering
\caption{Statistics for trials.
}
 \scalebox{0.84}{
\begin{tabular}{c|c|c||c|c} \\
&  \multicolumn{2}{c||}{\textit{LibriSpeech-train-960}} & \multicolumn{2}{c}{\textit{LibriSpeech-train-clean-360}} \\ \cline{2-5}
\textbf{\makecell{Average $\#$ utter. \\ per trial}} &   \makecell{Same \\ speaker} & \makecell{Different \\speaker} &   \makecell{Same \\speaker} & \makecell{Different \\ speaker}
   \\ \hline
1 &	\numprint{17527076}	& \multirow{7}{*}{\numprint{233800}} & \numprint{5944163} & \multirow{7}{*}{\numprint{92100}} 	 \\
3&	\numprint{1816610} &  & \numprint{716384} & 	\\
5&	\numprint{644790}	&  & \numprint{253935} & \\
10&	\numprint{154308}	&  & \numprint{60786} & 	\\
20&	\numprint{35070}	&  & \numprint{13815} & 	\\
40&	\numprint{7014} 	&  & \numprint{2763} & 	\\
60&	\numprint{2338} &  & \numprint{921} &	\\
\end{tabular}
}
\label{tab:trials}
\end{table}


\subsection{Anonymization systems}
\label{sec:anonym_systems}

To investigate the impact of speaker voice anonymization, we  consider two different state-of-the-art speaker  voice anonymization systems (SASs):
\begin{itemize}
    \item \textbf{SAS-1} keeps the original temporal phoneme dynamics, but changes the other speaker characteristics (speaker identity and some prosodical characteristics such as pitch and energy).
    \item \textbf{SAS-2} is a cascaded ASR-TTS system  that changes phoneme durations.
\end{itemize}
Below we briefly describe these systems.

\subsubsection{SAS-1}
\label{sec:sas1}

SAS-1, proposed in \cite{meyer2023prosody}
and used as baseline \textbf{B3} in the VoicePrivacy 2024 Challenge \cite{tomashenko2024voiceprivacy}, is a system based on
anonymization using phonetic transcription and a GAN that generates artificial pseudo-speaker embeddings. Anonymization is performed in three steps: (1) 
extraction of the speaker embedding, phonetic transcription, pitch,
energy, and phone duration from the original audio waveform; (2) speaker embedding anonymization, pitch and energy modification; and (3) synthesis of an anonymized speech waveform from the anonymized speaker embedding, modified pitch and energy features, original phonetic transcripts and original phone durations.

\subsubsection{SAS-2}
\label{sec:sas2}

SAS-2, proposed in \cite{jhu_2024_spsc}, is one of the best systems developed by the VoicePrivacy 2024 Challenge participants in terms of linguistic content and privacy preservation. It is a cascaded ASR-TTS system, where first the text transcripts are obtained from the source audio and then a TTS system is used to generate corresponding anonymized speech from the obtained transcripts with a new anonymized speaker voice. The ASR model is the
\textit{medium} English  \textit{Whisper} model \cite{radford2023robust}. The TTS model is \textit{VITS} (variational inference with adversarial
learning for end-to-end text-to-speech, \cite{kim2021conditional}), trained on the \textit{LibriTTS} dataset \cite{zen2019libritts}. 

\subsubsection{Automatic speaker verification and speech recognition results for SAS systems}
\label{sec:results_sas_base}

The ASV results in terms of equal error rate (EER) and the automatic speech recognition (ASR) results in terms of word error rate (WER) are shown in Table~\ref{tab:EER-base} on the \textit{LibriSpeech} test set for original and anonymized data. The trial lists for ASV evaluation in the \textit{LibriSpeech}  test data are taken from the VoicePrivacy 2024 Challenge \cite{tomashenko2024voiceprivacy} setup. ASV evaluation for original and anonymized data was performed with the same ASV model architecture and training setup as proposed in the VoicePrivacy 2024 Challenge. For anonymized data, the strongest \textit{semi-informed} attack models, trained on the \textit{utterance-level} anonymized data were used in evaluation.

\begin{table}[ht!]
\centering
\caption{EER (\%) and WER (\%) on original and anonymized data from the \textit{LibriSpeech test} dataset.}
 \scalebox{0.84}{
\begin{tabular}{c|c|c|c|} \\
\textbf{System} &  \textbf{EER,\% female} & \textbf{EER,\% male} & \textbf{WER,\%} \\ \hline
\textbf{Original} & 8.8 & 0.4 & 1.85 \\
\textbf{SAS-1} & 27.9 & 26.7 & 4.35 \\ 
\textbf{SAS-2} & 47.5 & 48.8 & 3.76 \\
\end{tabular}
}
\label{tab:EER-base}
\end{table}

\subsection{Results}
\label{sec:results}

To analyse speech temporal dynamics we performed several series of experiments dedicated to (1)  impact of the speaker's phoneme durations on the ASV performance and  metric comparison for ASV; (2) impact of the phoneme set on the  ASV performance with the proposed models; (3) impact of the speech rate on the ASV performance of the proposed attack models and the effect of normalizing all speakers to the same speech rate;
and (4) effect of different anonymization strategies on the ASV performance of the proposed systems.

\subsubsection{Speaker verification using phoneme durations and metric choice}

The ASV  results obtained using metric $\rho_1$ and $38$ phonemes are given in Table~\ref{tab:2-orig-960-p1-38} in terms of EER. Each line corresponds to the EERs obtained when the average number of utterances per speaker used to compute metric $\rho_1$  equals to the value in the first column ("Average $\#$ utter. per trial"). Different columns (1,3,\ldots, 20 -- minimum number of phoneme instances for averaging) correspond to  different values of the threshold parameter as defined in Section~\ref{sec:metrics}.
Increasing the number of utterances used to compute the speaker similarity metric allows us to significantly reduce the EER down to $9.2\%$ with 60 utterances per speaker.

The results reported in Table~\ref{tab:tab:3-orig-960-p2-38} for metric $\rho_2$ on the same data show similar trends. Also we can see that this metric is more efficient than $\rho_1$ when more utterances are used. Thus we use $\rho_2$ in the following experiments.

\newcommand{\g}[1]{\gradientcelld{#1}{2}{25}{50}{YellowOrange}{white}{ForestGreen}{70}}

\begin{table}[h]
\centering
\caption{EER (\%) on original data from the \textit{LibriSpeech-train-960} dataset obtained using metric $\rho_1$ and $N=38$ phonemes.}
 \scalebox{0.84}{
\begin{tabular}{c|c|c|c|c|c} \\
\multirow{2}{*}{\textbf{\makecell{Average $\#$ utter. \\ per trial}}} &   \multicolumn{5}{c}{\textbf{Minimum $\#$   phoneme instances for aver.}} \\
 &  1 & 3 & 5 & 10 & 20 \\ \hline
1 &	\g{39.9} & \g{38.5}	& \g{38.7} & \g{39.8} &	\g{40.3} \\
3 &	\g{34.9} & \g{32.4} & \g{32.0} & \g{32.2} & \g{34.9} \\
5 &	\g{31.9} & \g{28.4}	& \g{27.8} & \g{27.9} & \g{29.3} \\
10 & \g{28.2} &	\g{23.1} & \g{22.3}	& \g{22.3} & \g{23.4} \\
20 & \g{22.4} &	\g{23.4} & \g{18.1} & \g{16.7} & \g{17.1} \\
40 & \g{16.0} & \g{17.7} &	\g{21.0} & \g{12.8}	& \g{12.3} \\
60 & \g{12.8} &	\g{13.2} & \g{16.8} & \g{15.4} & \g{9.2} \\
\end{tabular}
}
\label{tab:2-orig-960-p1-38}
\end{table}

\begin{table}[h]
\centering
\caption{EER (\%) on original data from the \textit{LibriSpeech-train-960} dataset obtained using metric $\rho_2$ and $N=38$ phonemes.}
 \scalebox{0.84}{
\begin{tabular}{c|c|c|c|c|c} \\
\multirow{2}{*}{\textbf{\makecell{Average $\#$ utter. \\ per trial}}} &   \multicolumn{5}{c}{\textbf{Minimum $\#$   phoneme instances for aver.}} \\
 &  1 & 3 & 5 & 10 & 20 \\ \hline
1 &	\g{40.3} & \g{38.9}	& \g{39.4} & \g{39.9} &	\g{39.9} \\
3 &	\g{33.0} & \g{29.6} & \g{29.8} & \g{31.8} & \g{33.4} \\
5 &	\g{26.8} & \g{23.8}	& \g{22.6} & \g{23.8} & \g{27.3} \\
10 & \g{17.6} &	\g{16.5} & \g{15.1}	& \g{13.6} & \g{14.9} \\ 
20 & \g{10.2} &	\g{9.1} & \g{9.3} &	\g{7.7} & \g{6.9} \\
40 & \g{5.1} & \g{4.7} & \g{4.3} & \g{4.2}	& \g{3.2} \\
60 & \g{3.3} &	\g{3.1}	& \g{2.7} & \g{2.7} & \g{2.5} \\
\end{tabular}
}
\label{tab:tab:3-orig-960-p2-38}
\end{table}

\subsubsection{Selecting a set of acoustic units}

Table~\ref{tab:4-orig-960-p2-336} reports results with the increased number of phoneme classes: $N=336$.  Comparing them with Table~\ref{tab:tab:3-orig-960-p2-38}, we can see that increasing the number of phoneme classes does not provide improvement in EER.

\begin{table}[ht!]
\centering
\caption{EER (\%) on original data from the \textit{LibriSpeech-train-960} dataset obtained using metric $\rho_2$ and $N=336$ phoneme classes.}
 \scalebox{0.84}{
\begin{tabular}{c|c|c|c|c|c} \\
\multirow{2}{*}{\textbf{\makecell{Average $\#$ utter. \\ per trial}}} &   \multicolumn{5}{c}{\textbf{Minimum $\#$   phoneme instances for aver.}} \\
 &  1 & 3 & 5 & 10 & 20 \\ \hline
1 &	\g{39.3} & \g{39.7} & \g{39.9} & \g{39.8} &	\g{39.9} \\
3 &	\g{31.8} & \g{32.5} & \g{32.9} & \g{33.1} & \g{33.4} \\
5 &	\g{26.3} & \g{27.0} & \g{27.6} & \g{28.3} &	\g{28.8} \\
10 & \g{18.7} & \g{19.5} & \g{20.0}	& \g{21.1} & \g{22.3} \\
20 & \g{12.0} &	\g{12.5} & \g{13.1} & \g{14.0} &	\g{15.5} \\
40 & \g{6.5} & \g{6.6} & \g{6.5} & \g{7.7} & \g{8.7} \\
60 & \g{4.4} & \g{3.6} & \g{4.0} & \g{4.5} & \g{5.6} \\
\end{tabular}
}
\label{tab:4-orig-960-p2-336}
\end{table}

\subsubsection{Speech rate as a discriminative feature and normalization of speech temporal dynamics to speech rate}

Table~\ref{tab:5} shows ASV performance based on speaker's speech rate.
The speech rate was calculated as $\frac{\sum_{k=1}^K{\bar{l}_k}}{\sum_{k=1}^K{{l_k}}}$, where $K$ is the number of phones in the utterance, $l_k$ is the actual duration of  phone~$k$
in the utterance, $\bar{l}_k$ is the expected mean duration of the corresponding phoneme $k$ estimated from the training
corpus.
We can see that speech rate allows us to successfully perform ASV although, when the average number of utterances is larger than 3, the EER is higher in comparison with the cases when we use phoneme-based temporal characteristics. 


\begin{table}[ht!]
\centering
\caption{EER (\%) on original data from the \textit{LibriSpeech-train-960} dataset obtained using metric $\rho_2$ and speech rate.}
 \scalebox{0.84}{
\begin{tabular}{c|c} \\
\multirow{2}{*}{\textbf{\makecell{Average $\#$ utter. \\ per trial}}} &   \multicolumn{1}{c}{} \\
   \\ \hline
1 &	\g{38.6}	\\
3 &	\g{31.9}	\\
5 &	\g{27.4} 	\\
10 & \g{22.1}	\\
20 & \g{17.8}	\\
40 & \g{13.9} 	\\
60 & \g{11.8}   \\
\end{tabular}
}
\label{tab:5}
\end{table}

Tables~\ref{tab:6-orig-960-p2-38} and \ref{tab:7} show the ASV results after performing global speech rate normalization. In these experiments, we first computed phoneme duration statistics over the full \textit{LibriSpeech-train-960} corpus and then the speech rate of each utterance was adjusted with a constant factor to match the average speech rate.
As expected, such normalization degrades the performance of the ASV systems compared to the results without normalization in Tables~\ref{tab:tab:3-orig-960-p2-38} and \ref{tab:4-orig-960-p2-336} in most cases.
However, interestingly, normalization achieves lower EER results ($2\%$) when using a large number of utterances (60) and $N=336$. 


\begin{table}[ht!]
\centering
\caption{EER (\%) on original data from the \textit{LibriSpeech-train-960} dataset obtained using metric $\rho_2$ with global speech rate normalization and $N=38$ phonemes.}
 \scalebox{0.84}{
\begin{tabular}{c|c|c|c|c|c} \\
\multirow{2}{*}{\textbf{\makecell{Average $\#$ utter. \\ per trial}}} &   \multicolumn{5}{c}{\textbf{Minimum $\#$   phoneme instances for aver.}} \\
 &  1 & 3 & 5 & 10 & 20 \\ \hline
1 &	\g{45.1} & \g{45.5}	& \g{46.7} & \g{48.3} &	\g{49.0} \\
3 &	\g{38.8} & \g{36.5}	& \g{37.6} & \g{41.0} &	\g{44.0} \\
5 &	\g{33.3} & \g{30.6} & \g{29.8} & \g{32.0} & \g{37.2} \\
10 & \g{24.0} &	\g{22.8} & \g{21.0} & \g{19.9} & \g{21.8} \\
20 & \g{14.6} &	\g{13.7} & \g{13.6}	& \g{11.5} & \g{10.9} \\
40 & \g{7.3} & \g{7.2} & \g{6.7} & \g{6.5} & \g{4.8} \\
60 & \g{4.7} & \g{4.4} & \g{4.7} & \g{4.2} & \g{3.8} \\
\end{tabular}
}
\label{tab:6-orig-960-p2-38}
\end{table}

\begin{table}[ht!]
\centering
\caption{EER (\%) on original data from the \textit{LibriSpeech-train-960} dataset obtained using metric $\rho_2$ with global speech rate normalization and $N=336$ phoneme classes.}
 \scalebox{0.84}{
\begin{tabular}{c|c|c|c|c|c} \\
\multirow{2}{*}{\textbf{\makecell{Average $\#$ utter. \\ per trial}}} &   \multicolumn{5}{c}{\textbf{Minimum $\#$   phoneme instances for aver.}} \\
 &  1 & 3 & 5 & 10 & 20 \\ \hline
1 &	\g{46.7} & \g{47.8}	& \g{48.0}	& \g{48.1} & \g{48.2} \\
3 &	\g{40.2} & \g{42.6}	& \g{44.0}	& \g{45.1} & \g{45.7} \\
5 &	\g{33.7} & \g{35.7} & \g{38.0}	& \g{41.0} & \g{42.6} \\
10 & \g{23.9} &	\g{22.9} & \g{24.1} & \g{28.7} & \g{34.2} \\
20 & \g{15.4} &	\g{12.4} & \g{12.3}	& \g{13.3} & \g{18.1} \\
40 & \g{8.6} & \g{5.2} & \g{4.9} &	\g{4.8}	& \g{5.4} \\
60 & \g{5.6} & \g{3.1} & \g{2.4} &	\g{2.5} & \g{2.0} \\
\end{tabular}
}
\label{tab:7}
\end{table}

\begin{table}[ht!]
\centering
\caption{EER (\%) on original data from the \textit{LibriSpeech-train-clean-360} dataset obtained using metric $\rho_2$ and $N=38$ phonemes.}
 \scalebox{0.84}{
\begin{tabular}{c|c|c|c|c|c} \\
\multirow{2}{*}{\textbf{\makecell{Average $\#$ utter. \\ per trial}}} &   \multicolumn{5}{c}{\textbf{Minimum $\#$   phoneme instances for aver.}} \\
 &  1 & 3 & 5 & 10 & 20 \\ \hline
1 &	\g{40.4} & \g{39.2}	& \g{39.6} & \g{40.1} &	\g{40.1} \\
3 &	\g{34.7} & \g{31.7} & \g{31.7} & \g{33.4} & \g{34.3} \\
5 &	\g{28.1} & \g{25.6}	& \g{24.7} & \g{26.0} &	\g{28.9} \\
10 & \g{18.5} &	\g{17.6} & \g{16.1}	& \g{15.5} & \g{16.9} \\
20 & \g{10.4} &	\g{9.6} & \g{9.7} &	\g{8.2} & \g{8.0} \\
40 & \g{4.9} & \g{4.6} & \g{4.1} & \g{3.9}	& \g{3.7} \\
60 & \g{2.7} & \g{2.9} & \g{3.0} & \g{2.5} & \g{2.5} \\
\end{tabular}
}
\label{tab:8}
\end{table}

\subsubsection{Experiments on anonymized data}

Experiments on anonymized \textit{LibriSpeech-train-clean-360} data for the two anonymization systems SAS-1 and SAS-2 are reported in Tables~\ref{tab:9} and \ref{tab:10}, respectively.  For comparison purposes, we also added results on the original data for the same dataset in Table~\ref{tab:8}.
SAS-1 does not change phoneme durations and we can see despite some degradation of results (in Table~\ref{tab:9} vs.\ \ref{tab:8}) that the preserved speech dynamics still allow us to retrieve speaker information (the lowest EER is $7\%$). SAS-2 changes phoneme durations and as expected provides much higher privacy protection (Table~\ref{tab:10}). However, surprisingly, for a large number of utterances (60), the EER is still low ($26.3\%$). 
One possible explanation might be that, in read speech, the book content may impact the speaking style and thus  temporal dynamic statistics.

\begin{table}[ht!]
\centering
\caption{EER (\%) on data anonymized by SAS-1 from the \textit{LibriSpeech-train-clean-360} dataset obtained using metric $\rho_2$ and $N=38$ phonemes.}
 \scalebox{0.84}{
\begin{tabular}{c|c|c|c|c|c} \\
\multirow{2}{*}{\textbf{\makecell{Average $\#$ utter. \\ per trial}}} &   \multicolumn{5}{c}{\textbf{Minimum $\#$   phoneme instances for aver.}} \\
 &  1 & 3 & 5 & 10 & 20 \\ \hline
1 &	\g{42.3} & \g{40.2}	& \g{40.2} & \g{40.4} &	\g{40.3} \\
3 &	\g{37.9} & \g{35.1} & \g{34.2} & \g{34.3} & \g{34.7} \\
5 &	\g{33.2} & \g{30.5}	& \g{29.3} & \g{28.9} & \g{29.8} \\
10 & \g{25.4} &	\g{24.1} & \g{23.1}	& \g{21.3} & \g{21.3} \\
20 & \g{17.7} &	\g{16.7} & \g{16.6} & \g{15.4} & \g{13.9} \\
40 & \g{10.4} & \g{10.2} & \g{9.8} & \g{9.6} & \g{8.5} \\
60 & \g{7.4} & \g{7.3} & \g{7.2} & \g{7.0} & \g{7.1} \\
\end{tabular}
}
\label{tab:9}
\end{table}

\begin{table}[ht!]
\centering
\caption{EER (\%) on data anonymized by SAS-2 from the \textit{LibriSpeech-train-clean-360} dataset obtained using metric $\rho_2$ and $N=38$ phonemes.}
 \scalebox{0.84}{
\begin{tabular}{c|c|c|c|c|c} \\
\multirow{2}{*}{\textbf{\makecell{Average $\#$ utter. \\ per trial}}} &   \multicolumn{5}{c}{\textbf{Minimum $\#$   phoneme instances for aver.}} \\
 &  1 & 3 & 5 & 10 & 20 \\ \hline
1 &	\g{49.0} & \g{49.4}	& \g{49.4} & \g{49.3} & \g{49.7} \\
3 &	\g{47.7} & \g{47.4} & \g{47.5} & \g{48.4} & \g{48.7} \\
5 &	\g{46.3}  & \g{45.6} & \g{45.6} & \g{46.0} & \g{47.8} \\
10 & \g{43.4} &	\g{43.1} & \g{42.1}	& \g{41.8} & \g{41.9} \\
20 & \g{39.1} &	\g{38.6} & \g{38.7}	& \g{36.2} & \g{36.8} \\
40 & \g{32.1} &	\g{32.0} & \g{31.4} & \g{31.5} & \g{28.0} \\
60 & \g{27.6} & \g{27.0} & \g{26.3} & \g{27.5} & \g{26.3} \\
\end{tabular}
}
\label{tab:10}
\end{table}

\section{Conclusions}
\label{sec:prior}
In this study, we demonstrated the importance of speech temporal dynamics analysis which has been under-explored in voice anonymization research to date.
Using the proposed metrics and sufficient amount of data per speaker, we achieve an EER as low as $7\%$ on the anonymized data obtained by a speaker voice anonymization system that does not modify phoneme durations.
In future work, we plan to verify
the observed phenomena on other types of speech data, in particular on spontaneous speech, and to improve state-of-the-art anonymization techniques by integrating temporal dynamics normalization. 
The proposed simple approach to analyze temporal dynamics shows the potential for more advanced analysis by means of machine learning (ML) models that will allow integrating multiple discovered discriminative factors into ML models and performing more fine-grained and efficient analysis, e.g., using attention mechanisms.

\bibliographystyle{IEEEbib}
\bibliography{refs}

\begin{thebibliography}{10}

\bibitem{tomashenko2020introducing}
Natalia Tomashenko, Brij Mohan~Lal Srivastava, Xin Wang, Emmanuel Vincent, Andreas Nautsch, Junichi Yamagishi, Nicholas Evans, Jose Patino, Jean-François Bonastre, Paul-Gauthier Noé, and Massimiliano Todisco,
\newblock ``{Introducing the {VoicePrivacy} Initiative},''
\newblock in {\em Interspeech}, 2020, pp. 1693--1697.

\bibitem{patino2020speaker}
Jose Patino, Natalia Tomashenko, Massimiliano Todisco, Andreas Nautsch, and Nicholas Evans,
\newblock ``Speaker anonymisation using the {McAdams} coefficient,''
\newblock in {\em Interspeech}, 2021, pp. 1099--1103.

\bibitem{mawalim22_spsc}
Candy~Olivia Mawalim, Shogo Okada, and Masashi Unoki,
\newblock ``Speaker anonymization by pitch shifting based on time-scale modification,''
\newblock in {\em 2nd Symposium on Security and Privacy in Speech Communication}, 2022, pp. 35--42.

\bibitem{gupta2020designn}
Priyanka Gupta, Gauri~P. Prajapati, Shrishti Singh, Madhu~R. Kamble, and Hemant~A. Patil,
\newblock ``Design of voice privacy system using linear prediction,''
\newblock in {\em 2020 Asia-Pacific Signal and Information Processing Association Annual Summit and Conference (APSIPA ASC)}, 2020, pp. 543--549.

\bibitem{tavi2022improving}
Lauri Tavi, Tomi Kinnunen, and Rosa~Gonz{\'a}lez Hautam{\"a}ki,
\newblock ``Improving speaker de-identification with functional data analysis of f0 trajectories,''
\newblock {\em Speech Communication}, vol. 140, pp. 1--10, 2022.

\bibitem{srivastava2021}
Brij Mohan~Lal Srivastava, Mohamed Maouche, Md~Sahidullah, Emmanuel Vincent, Aurélien Bellet, Marc Tommasi, Natalia Tomashenko, Xin Wang, and Junichi Yamagishi,
\newblock ``Privacy and utility of x-vector based speaker anonymization,''
\newblock {\em IEEE/ACM Transactions on Audio, Speech and Language Processing}, vol. 30, pp. 2383--2395, 2022.

\bibitem{miao2023speaker}
Xiaoxiao Miao, Xin Wang, Erica Cooper, Junichi Yamagishi, and Natalia Tomashenko,
\newblock ``Speaker anonymization using orthogonal {Householder} neural network,''
\newblock {\em IEEE/ACM Transactions on Audio, Speech, and Language Processing}, vol. 31, pp. 3681--3695, 2023.

\bibitem{champion2023}
Pierre Champion,
\newblock {\em Anonymizing speech: evaluating and designing speaker anonymization techniques},
\newblock Ph.D. thesis, Université de Lorraine, 2023.

\bibitem{yao2024musa}
Jixun Yao, Qing Wang, Pengcheng Guo, Ziqian Ning, Yuguang Yang, Yu~Pan, and Lei Xie,
\newblock ``{MUSA}: Multi-lingual speaker anonymization via serial disentanglement,''
\newblock {\em arXiv preprint arXiv:2407.11629}, 2024.

\bibitem{tomashenko2024voiceprivacy}
Natalia Tomashenko, Xiaoxiao Miao, Pierre Champion, Sarina Meyer, Xin Wang, Emmanuel Vincent, et~al.,
\newblock ``The {VoicePrivacy} 2024 challenge evaluation plan,''
\newblock {\em arXiv preprint arXiv:2404.02677}, 2024.

\bibitem{fang2019speaker}
Fuming Fang, Xin Wang, Junichi Yamagishi, Isao Echizen, Massimiliano Todisco, Nicholas Evans, and Jean-Francois Bonastre,
\newblock ``Speaker anonymization using x-vector and neural waveform models,''
\newblock in {\em Speech Synthesis Workshop}, 2019, pp. 155--160.

\bibitem{meyer2023prosody}
Sarina Meyer, Florian Lux, Julia Koch, Pavel Denisov, Pascal Tilli, and Ngoc~Thang Vu,
\newblock ``Prosody is not identity: A speaker anonymization approach using prosody cloning,''
\newblock in {\em IEEE International Conference on Acoustics, Speech and Signal Processing (ICASSP)}, 2023, pp. 1--5.

\bibitem{panariello_speaker_2023}
Michele Panariello, Francesco Nespoli, Massimiliano Todisco, and Nicholas Evans,
\newblock ``Speaker anonymization using neural audio codec language models,''
\newblock in {\em IEEE International Conference on Acoustics, Speech and Signal Processing (ICASSP)}, 2024, pp. 4725--4729.

\bibitem{miao2022language}
Xiaoxiao Miao, Xin Wang, Erica Cooper, Junichi Yamagishi, and N~Tomashenko,
\newblock ``Language-independent speaker anonymization approach using self-supervised pre-trained models,''
\newblock {\em arXiv preprint arXiv:2202.13097}, 2022.

\bibitem{sinha2022eli}
Yamini Sinha, Jan Hintz, Matthias Busch, Tim Polzehl, Matthias Haase, Andreas Wendemuth, and Ingo Siegert,
\newblock ``Why {Eli Roth} should not use {TTS-systems} for anonymization,''
\newblock in {\em Proceedings of the 2nd Symposium on Security and Privacy in Speech Communication}, 2022, pp. 17--22.

\bibitem{prajapati2022voice}
Gauri~P. Prajapati, Dipesh~K. Singh, Preet~P. Amin, and Hemant~A. Patil,
\newblock ``Voice privacy using {CycleGAN} and time-scale modification,''
\newblock {\em Computer Speech \& Language}, vol. 74, 2022.

\bibitem{Tomashenko2021CSl}
Natalia Tomashenko, Xin Wang, Emmanuel Vincent, Jose Patino, Brij Mohan~Lal Srivastava, Paul-Gauthier Noé, Andreas Nautsch, Nicholas Evans, Junichi Yamagishi, et~al.,
\newblock ``{The VoicePrivacy 2020 Challenge: Results and findings},''
\newblock {\em Computer Speech and Language}, vol. 74, pp. 101362, 2022.

\bibitem{tomashenko2024first}
Natalia Tomashenko, Xiaoxiao Miao, Emmanuel Vincent, and Junichi Yamagishi,
\newblock ``The first {VoicePrivacy Attacker Challenge} evaluation plan,''
\newblock {\em arXiv preprint arXiv:2410.07428}, 2024.

\bibitem{desplanques2020ecapa}
Brecht Desplanques, Jenthe Thienpondt, and Kris Demuynck,
\newblock ``{ECAPA-TDNN}: Emphasized channel attention, propagation and aggregation in {TDNN} based speaker verification,''
\newblock in {\em Interspeech}, 2020, pp. 3830--3834.

\bibitem{bulgakova2015speaker}
Elena Bulgakova, Aleksei Sholohov, Natalia Tomashenko, and Yuri Matveev,
\newblock ``Speaker verification using spectral and durational segmental characteristics,''
\newblock in {\em 17th International Conference on Speech and Computer}, 2015, pp. 397--404.

\bibitem{fujita21_interspeech}
Kenichi Fujita, Atsushi Ando, and Yusuke Ijima,
\newblock ``Phoneme duration modeling using speech rhythm-based speaker embeddings for multi-speaker speech synthesis,''
\newblock in {\em Interspeech}, 2021, pp. 3141--3145.

\bibitem{fujita2024speech}
Kenichi Fujita, Atsushi Ando, and Yusuke Ijima,
\newblock ``Speech rhythm-based speaker embeddings extraction from phonemes and phoneme duration for multi-speaker speech synthesis,''
\newblock {\em IEICE Transcations on Information and Systems}, vol. 107, no. 1, pp. 93--104, 2024.

\bibitem{panayotov2015librispeech}
Vassil Panayotov, Guoguo Chen, Daniel Povey, and Sanjeev Khudanpur,
\newblock ``{LibriSpeech}: an {ASR} corpus based on public domain audio books,''
\newblock in {\em IEEE International Conference on Acoustics, Speech and Signal Processing (ICASSP)}, 2015, pp. 5206--5210.

\bibitem{povey2011kaldi}
Daniel Povey, Arnab Ghoshal, Gilles Boulianne, Lukas Burget, Ondrej Glembek, Nagendra Goel, Mirko Hannemann, Petr Motl\'i\v{c}ek, et~al.,
\newblock ``The {Kaldi} speech recognition toolkit,''
\newblock in {\em IEEE Automatic Speech Recognition and Understanding Workshop (ASRU)}, 2011.

\bibitem{jhu_2024_spsc}
Henry~Li Xinyuan, Zexin Cai, Ashi Garg, Kevin Duh, Leibny~Paola García-Perera, Sanjeev Khudanpur, Nicholas Andrews, and Matthew Wiesn,
\newblock ``{HLTCOE JHU} submission to the {Voice Privacy} challenge,''
\newblock in {\em 4th Symposium on Security and Privacy in Speech Communication}, 2024.

\bibitem{radford2023robust}
Alec Radford, Jong~Wook Kim, Tao Xu, Greg Brockman, Christine McLeavey, and Ilya Sutskever,
\newblock ``Robust speech recognition via large-scale weak supervision,''
\newblock in {\em International Conference on Machine Learning (ICML)}, 2023.

\bibitem{kim2021conditional}
Jaehyeon Kim, Jungil Kong, and Juhee Son,
\newblock ``Conditional variational autoencoder with adversarial learning for end-to-end text-to-speech,''
\newblock in {\em International Conference on Machine Learning (ICML)}, 2021, pp. 5530--5540.

\bibitem{zen2019libritts}
Heiga Zen, Viet Dang, Rob Clark, Yu~Zhang, Ron~J Weiss, Ye~Jia, Zhifeng Chen, and Yonghui Wu,
\newblock ``{LibriTTS}: A corpus derived from {LibriSpeech} for text-to-speech,''
\newblock in {\em Interspeech}, 2019, pp. 1526--1530.

\end{thebibliography}

\end{document}